\documentstyle[11pt,epsfig]{article}
\newfont{\Bbb}{msbm10 scaled 1200}     
\newcommand{\mathbb}[1]{\mbox{\Bbb #1}}

\def\lbldef#1#2{\expandafter\gdef\csname #1\endcsname {#2}}

\def\href#1#2{#2}

\newcommand{\beq}{\begin{equation}}
\newcommand{\eeq}{\end{equation}}

\newcommand{\ber}{\begin{eqnarray}}
\newcommand{\eer}{\end{eqnarray}}

\newcommand{\beqar}{\begin{eqnarray}}

\newcommand{\eeqar}{\end{eqnarray}}

\newcommand{\ba}{\begin{eqnarray}}
\newcommand{\ea}{\end{eqnarray}}

\newcommand{\dsl}
  {\kern.06em\hbox{\raise.15ex\hbox{$/$}\kern-.56em\hbox{$\partial$}}}

\newcommand{\eeqarr}{\end{eqnarray}}
\newcommand{\ZZ}{{\rm \kern 0.275em Z \kern -0.92em Z}\;}

\def\be{\begin{equation}}
\def\ee{\end{equation}}
\def\bea{\begin{eqnarray}}
\def\eea{\end{eqnarray}}

\makeatletter
\def\flE{\begin{picture}(0,0)
   \put( 0.25,    0){\vector( 1, 0){0.50}}
   \@ifstar{\@flE}{\@@flE}}
\def\@flE  #1{\put( 0.5 ,-0.03){\makebox(0,0)[ t]{$#1$}}\end{picture}}
\def\@@flE #1{\put( 0.5 , 0.03){\makebox(0,0)[ b]{$#1$}}\end{picture}}
\def\flNE{\begin{picture}(0,0)
   \put( 0.18, 0.18){\vector( 1, 1){0.64}}
   \@ifstar{\@flNE}{\@@flNE}}
\def\@flNE #1{\put( 0.52, 0.48){\makebox(0,0)[tl]{$#1$}}\end{picture}}
\def\@@flNE#1{\put( 0.48, 0.52){\makebox(0,0)[br]{$#1$}}\end{picture}}
\def\flN{\begin{picture}(0,0)
   \put(    0, 0.20){\vector( 0, 1){0.60}}
   \@ifstar{\@flN}{\@@flN}}
\def\@flN  #1{\put( 0.03, 0.5 ){\makebox(0,0)[ l]{$#1$}}\end{picture}}
\def\@@flN #1{\put(-0.03, 0.5 ){\makebox(0,0)[ r]{$#1$}}\end{picture}}
\def\flNW{\begin{picture}(0,0)
   \put(-0.18, 0.18){\vector(-1, 1){0.64}}
   \@ifstar{\@flNW}{\@@flNW}}
\def\@flNW #1{\put(-0.48, 0.52){\makebox(0,0)[bl]{$#1$}}\end{picture}}
\def\@@flNW#1{\put(-0.52, 0.48){\makebox(0,0)[tr]{$#1$}}\end{picture}}
\def\flW{\begin{picture}(0,0)
   \put(-0.25,    0){\vector(-1, 0){0.50}}
   \@ifstar{\@flW}{\@@flW}}
\def\@flW  #1{\put(-0.5 , 0.03){\makebox(0,0)[ b]{$#1$}}\end{picture}}
\def\@@flW #1{\put(-0.5 ,-0.03){\makebox(0,0)[ t]{$#1$}}\end{picture}}
\def\flSW{\begin{picture}(0,0)
   \put(-0.18,-0.18){\vector(-1,-1){0.64}}
   \@ifstar{\@flSW}{\@@flSW}}
\def\@flSW #1{\put(-0.52,-0.48){\makebox(0,0)[br]{$#1$}}\end{picture}}
\def\@@flSW#1{\put(-0.48,-0.52){\makebox(0,0)[tl]{$#1$}}\end{picture}}
\def\flS{\begin{picture}(0,0)
   \put(    0,-0.2 ){\vector( 0,-1){0.60}}
   \@ifstar{\@flS}{\@@flS}}
\def\@flS  #1{\put(-0.03,-0.5 ){\makebox(0,0)[ r]{$#1$}}\end{picture}}
\def\@@flS #1{\put( 0.03,-0.5 ){\makebox(0,0)[ l]{$#1$}}\end{picture}}
\def\flSE{\begin{picture}(0,0)
   \put( 0.18,-0.18){\vector( 1,-1){0.64}}
   \@ifstar{\@flSE}{\@@flSE}}
\def\@flSE #1{\put( 0.48,-0.52){\makebox(0,0)[tr]{$#1$}}\end{picture}}
\def\@@flSE#1{\put( 0.52,-0.48){\makebox(0,0)[bl]{$#1$}}\end{picture}}
\def\capsa(#1,#2)#3{\put(#1,#2){\makebox(0,0){$#3$}}}
\def\indiag{\@ifnextchar [{\@indiag}{\@indiag[15ex]}}
\def\@indiag[#1](#2,#3){\begingroup
   \setlength{\unitlength}{#1}
   \medskip
   \begin{center}
   \begin{picture}(#2,#3)}
\def\exdiag{\end{picture}
   \end{center}
   \medskip
   \endgroup}
\makeatother

\begin{document}
\baselineskip=15.5pt
\pagestyle{plain}
\setcounter{page}{1}
\begin{titlepage}


\vskip -.8cm


\begin{center}

\vskip 1.7 cm

{\LARGE \bf Dimensional reduction, truncations, constraints \\ and the issue of
consistency\footnote{Talk presented at the NEB XII conference 
 Recent Developments in Gravity, June 29th - July 2nd 2006, Nafplion, Greece. }}
\vskip .3cm

\vskip 1.5cm
\vspace{20pt}
{\large
J.M. Pons 
}
\vskip 1.2cm

\textit{
Departament ECM, Facultat de F\'isica, Universitat de Barcelona and \\ Institut
de Fisica d'Altes Energies, Diagonal 647, E-08028 Barcelona, Spain.}\\

\vskip 0.5cm

\vspace{1cm}

{\bf Abstract}

\end{center}

A brief overview of dimensional reductions 
for diffeomorphism invariant theories is given. The distinction between the 
physical idea of compactification and the mathematical problem of a consistent truncation 
is discussed, and the typical ingredients of the latter --reduction of spacetime 
dimensions and the introduction of constraints-- are examined. 
The consistency in the case of of group manifold reductions, when the structure constants 
satisfy the unimo\-dularity condition, is shown together 
with the associated reduction of the gauge group. 
The problem of consistent truncations on coset spaces
is also discussed and we comment on examples of some remarkable consistent 
truncations that have been found in this context.     

\noindent

\end{titlepage}

\newpage


\section{Introduction. Some brief highlights}
\label{intro}

Itr seems that the first reference to dimensional reduction 
(see \cite{Duff:hr} and \cite{Appelquist:nr}
as general references) appears in the
work of Gunnar Nordstr\"om \cite{gn} who in 1914 formulated a vector-scalar theory 
in four dimensions --unifying electromagnetism and a scalar theory of
gravitation-- starting from Maxwell theory in a five-dimensional flat spacetime. 
Much more known is the work of Theodor Kaluza \cite{kaluza}, published in 1921, who showed that
gravity in five dimensions could yield a unified gravity plus Maxwell 
(plus a scalar which was at that time ignored). It was Oskar Klein \cite{klein} who came with the idea of 
compactifying the fifth dimension on a circle. The --Fourier-- expansion in modes, today known as
the Kaluza-Klein tower, allowed him to compute the radius of compactification by
identifing the charge of the first massive mode with the electric charge. With a single stroke, 
the quantization of electric charge was given an explanation, and the size of the compact dimension, which turned
to be of the order of the Plank length, 
explained why we only see see effectively four dimensions. The other side of the story is the 
very wrong result for a mass of such a a mode, which was of the order of the Plank mass\footnote{
See \cite{Duff:1994tn} for a nice account.}.

A compactification on a 2-sphere, from six dimensions, was considered by Wolfgang Pauli 
in 1953 (see
\cite{pauli53}), were the Yang-Mills field strenght made its appearance as a consequence of the 
non-Abelian reduction. It was Bryce DeWitt \cite{dewitt;63} in 1963 who showed in full generality 
--in fact it was left in his Les Houches lectures as an exercise-- the unification of gravity and Yang Mills theories when
dimensionally reducing from gravity in higher dimensions.

The consistency problems of dimensional reduction where first raised by
Stephen Hawking \cite{hawking;69} in 1969 in the
context of the study of Bianchi cosmologies by dimensionally
reducing the Lagrangian of pure 4-dimensional GR under a three dimensional Lie
group. He found that when the trace of the structure constants had some
non-vanishing component (Bianchi's type B models), there was a
mismatch between the reduction of the original equations of
motion and the new equations of motion derived from the reduced
Lagrangian. This fact had been already noticed a little earlier by
Sch\"ucking, but never published \cite{bianchi}. In a more general
framework, the tracelessness of the structure constants as a
necessary condition for the consistency of dimensional reductions was formally
pointed out by Scherk and Schwarz in 1979, \cite{Scherk:1979zr},
although it has since remained almost forgotten and its interpretation somewhat 
obscure (in words of the authors of \cite{AlonsoAlberca:2003jq}, this condition is 
"a subtlety which is not obvious from the analysis by Scherk and Schwarz").

A different explanation of the tracelessness condition --also known as the 
unimodularity condition because it says that the adjoint representation is unimodular--, 
again as a necessary one, was given in the same year by MacCallum
\cite{Maccallum:gd} in the context of the Bianchi models, by
examining the boundary conditions required for the correct
application of the variational principle for solutions exhibiting
certain Killing symmetries.


Let us point out the two different philosophies that underlie the issue of dimensional 
reduction. On one side there is the {\it compactification} approach, whose
paramount example today is string theory, where one expects that
dimensional reduction from ten dimensions
---or eleven in M-theory---, will eventually deliver a four dimensional
space-time together with an internal compact manifold ---or
a more general structure like an orbifold--- that carries physical
information within. In this approach the fundamental theory is the
higher dimensional one, and the compactification procedure should
eventually be understood as a physical process driven by some
physical principles. The physical
consequences that have their origin in the compactified structure
and the process of compactification itself, are an essential part
of the whole physical picture. Perhaps
some of these effects, take for instance the presence of massive
Kaluza-Klein modes, may become irrelevant when the theory is
examined in the low energy regime, due to the small size of the
compactified structure, but the effects are there anyway. 
This was in fact Klein's elaboration
\cite{klein} (expanding in Fourier modes) on the original Kaluza
\cite{kaluza} idea, which corresponds to the second approach which we adress now.

On the other side one can conceive the dimensional reduction as a
mathematical means to formulate a theory in a given space-time
dimension having started with a higher dimensional theory as an
auxiliary artifact, the extra dimensions never physically
existing. In this case we speak of a {\it consistent truncation} 
form the higher dimensional theory to the lower one. One can
start with a higher dimensional theory whose formulation is
perhaps simpler and eventually end up with a more complicated
theory at a lower dimension with the remarkable advantage of
keeping full control of its symmetries (including supersymmetry
when appropriate) or even with the right internal symmetries one
was willing to implement. In Klein's intepretation of Kaluza's
work, this means that a truncation is made on the tower of Fourier
modes (or their generalizations to non-Abelian groups) to keep
only the singlet ones.  In this sense,
this second approach can be understood as a method of model
building, having used the higher dimensional theory as an
intermediate device (to use the language of
\cite{Appelquist:1983vs}), to be disposed of at the end, that
helps to formulate a fundamental theory at a lower dimension.
This has proven a useful method in
supergravity: in this way, different dimensional reductions of the
eleven dimensional supergravity (or the ten dimensional Type IIB)
theory have yielded a fairly good recollection of supergravity
theories with extended supersymmetry at lower dimensions.

\vspace{4mm}

In this contribution we will define what is a consistent truncation 
(Section 2) and classify it in two types. Next  
we will show (Section 3) the emergence of the tracelessness condition as
a necessary and sufficient condition for such a consistency in the case of group 
manifolds. 
In Section 4 we consider the introduction of constraints to further reduce the 
degrees of freedom and a natural contact is made with the Dirac-Bergmann theory 
of constrained systems. Finally
In Section 5 the difficulties associated to coset reductions are considered 
and some examples of 
consistent truncations are given.

\section{Truncations}

Consider a Lagrangian density 
${\mathcal L}$ as the starting point, for a certain number of
dimensions of the space-time. We can produce a truncation of it, yielding 
a reduced Lagrangian ${\mathcal L}_R$, by
essentially two methods --or a mixture of both:
\begin{itemize}
\item[{\it{i)}}] First-type: by reducing the dimension of the
space-time (Kaluza-Klein dimensional reduction)
while keeping unchanged the number of degrees of freedom attached
to every space-time point.
\end{itemize}
\begin{itemize}
\item[{\it{ii})}] Second-type: by introducing constraints that
reduce the number of independent fields --or field components--
defining the theory.
\end{itemize}
 In both
cases we are producing a {\sl truncation} in the field content of
the theory At this point an
issue of {\sl consistency} of such a truncation arises. 
Namely, whether the solutions of the
equations of motion (e.o.m.) for the truncated theory --with Lagrangian ${\mathcal L}_R$-- 
are still
solutions of the e.o.m. for the original ${\mathcal L}$. This
property is expressed graphically as the commutativity of the
following diagram, 
\indiag(1,1) \capsa(0,0){
\frac{\delta{\mathcal L}}{\delta \Phi} } \capsa(0,1){{\mathcal L}}
\capsa(0,1){\flS{{\rm e.o.m.}}} \capsa(0,0){\flE{{\rm Red}.}}
\capsa(0,1){\qquad\quad\flE{{\rm Red}.}} \capsa(1,0){\qquad\
\qquad \qquad\big(\frac{\delta{\mathcal L}}{\delta \Phi}
\big)_{\!_{\!R}}\!=0 \Leftrightarrow \frac{\delta{\mathcal
L}_{\!R}}{\delta \Phi}=0 } \capsa(1.5,1){{\mathcal L}_R}
\capsa(1.5,1){\flS{{\rm e.o.m.}}} \exdiag

\vspace{4mm}

\noindent
 
A proper definition is the following: {\sl A truncation is said to
be consistent when its implementation at the level of the
variational principle agrees with that at the level of the
equations of motion}, i.e., if both operations commute: first
truncate the Lagrangian and then obtain the equations of motion
(e.o.m.), or first obtain the equations of motion and then
truncate them. 

This variational principle perspective of a consistent truncation has been studied in the 
mathematical literature under the name of {\sl Principle of Symmetric Criticallity} in 
\cite{Palais1979} and it has been applied to general relativity in \cite{Fels:2001rv}
(see also \cite{Anderson:1999cn}).

Another, weaker concept of
consistent truncation, \cite{pope}, just proceeds through the e.o.m.:
one starts with a Lagrangian density and introduces some ansatz
for the reduction of the fields, which is then plugged into the
e.o.m.. If the original e.o.m. are compatible with such an ansatz,
the reduced e.o.m. will be considered as a consistent truncation
of the former ones. Unless said otherwise, we will use the terminology of 
consistent truncations in the strong, first sense.

To the process of truncation {\sl of theories},
${\mathcal L} \rightarrow{\mathcal L}_R$, going from top to
bottom, there corresponds an opposite process of uplifting {\sl of
solutions}, from bottom to top. The basic result in this respect is that {\sl a
consistent truncation guarantees that any solution of the
${\mathcal L}_R$ dynamics can be uplifted to a solution of the
${\mathcal L}$ dynamics}.

\section{First-type truncations}
\subsection{The unimodularity condition}
Here we will answer the following question in the framework of diffeomorphism 
invariant theories: given the set of solutions of a theory 
--with Lagrangian density ${\mathcal L}$--
that share the same algebra of Killing symmetries --which leave each of these solutions 
invariant--, is 
there a reduced variational principle describing exactly such a set? We will prove that the 
answer is in the positive  \cite{Pons:1998tt,Pons:2003ka} al least in the case when the 
Killing vectors are all independent and can
be written in a certain set of coordinates as ${\bf K}_{a}= K_a\!^{b}(y) \,  {\partial}_{y^b}$,
where $y^a$ are the coodinates along the orbits 
(which disappear under reduction ) and we have assumed that the components 
$K_a\!^{b}$ do not depend on the transversal coodinates $x^\mu$ (which survive the reduction).
These Killing vectors are independent generators of left action of a group, forming a Lie algebra
$$\displaystyle 
    [{\bf K}_{a},{\bf K}_{b}] = C^{c}_{ab}{\bf K}_{c} \,.
$$

Associated with the left action of the group, there are 
left-invariant vectors, $\displaystyle {\bf Y}_b= Y_b\!^{c}(y) \,  {\partial}_{y^c}$,
which generate a right action of the group:
$$\displaystyle 
    {\mathcal L}_{{\bf K}_a}{\bf Y}_b =
    \left[{\bf K}_a,{\bf Y}_b\right] = 0\,,\quad 
[{\bf Y}_{a},{\bf Y}_{b}] = - C^{c}_{ab}{\bf Y}_{c}\,.
$$

One can define the dual forms:\ $\displaystyle {\omega}^{a} = \omega^a_{b}(y) {\bf d}
y^{b}\,,\ \  {\omega}^{a}\cdot{\bf Y}_b=\delta^a_b,\  $ which satisfy
$$\displaystyle {\bf d} {\omega}^{a}
    = \frac{1}{2} C^{a}_{bc}\,{\omega}^{b}\! \wedge  {\omega}^{c}\,,\quad 
({\mathcal L}_{{\bf K}_a} {\omega}^{b}) = 0\,.  $$

Let us use the basis of forms $\displaystyle\{{\bf d}x^\mu, {\omega}^{a}\}$
to express our objects. The metric for instance, will be written as 
$$\displaystyle {\bf g} =  g_{\mu\nu} {\bf d}x^\mu {\bf d}x^\nu
        + g_{ab}\left(A^a_\mu {\bf d} x^\mu  +  {\omega}^{a}\right)
            \left(A^b_\nu {\bf d} x^\nu +  {\omega}^{b}\right)\,,$$
which is just a way to express the degrees of freedom associated to the metric: 
$$\displaystyle g_{\mu\nu}(x,y),\ g_{ab}(x,y),\ A^a_\mu(x,y)\,.$$
Now apply the Killing conditions. The $y$-dependences will be eliminated,
$$\displaystyle{\mathcal L}_{{\bf K}_c}{\bf g}=0 \
\Rightarrow \ 
 g_{\mu\nu}(x),\ g_{ab}(x),\  A^a_\mu(x)\,.$$

Notice that \ $\displaystyle\det{g} = (\det{g_{\mu\nu}})(\det{g_{ab}}) \vert\omega\vert^2\,.$
If we express the Lagrangian in the new basis
$$\displaystyle{\mathcal L} = 
{\mathcal L}\left(\Phi, \Phi_{\mu},\Phi_{\mu\nu},{\bf Y}_b(\Phi),
{\bf Y}_a{\bf Y}_b(\Phi)\right)=: \vert\omega\vert \tilde{\mathcal L}\,,$$
where $\Phi(x,y)$ is a generic component of a generic field. Then we can  
define the reduced Lagrangian
\beq\displaystyle{\mathcal L}_R(\Phi, \partial_\mu\Phi,\partial_{\mu\nu}\Phi )
    := \tilde{\mathcal L}(\Phi,
    \partial_\mu\Phi,\partial_{\mu\nu}\Phi,
        \,{\bf Y}_a\Phi =0,\, {\bf Y}_a{\bf Y}_b\Phi = 0)\,,
\label{redlagr}
\eeq
and the Euler-Lagrange derivatives become
$$\displaystyle
 \frac{\delta{\mathcal L}}{\delta \Phi}  =  \vert\omega\vert
 \left(\frac{\partial\tilde{\mathcal L}}{\partial \Phi}
 -\partial_{\mu}\frac{\partial\tilde{\mathcal L}}{\partial\Phi_\mu}+\frac{1}{2}
\partial_{\mu\nu}\frac{\partial\tilde{\mathcal
L}}{\partial\Phi_{\mu\nu}}  \right.$$
$$\displaystyle\ -({\bf Y}_a + C_{ac}^c)
               (\frac{\partial\tilde{\mathcal L}}{\partial{\bf
               Y}_a\Phi})
      + \frac{1}{2} \left.
         ({\bf Y}_b+C_{bd}^d)({\bf Y}_a + C_{ac}^c)
          (\frac{\partial\tilde{\mathcal L}}{\partial{\bf Y}_a{\bf Y}_b \Phi})
         \right)\,,$$
where crucial use has been made of the relation \ $\displaystyle
 \partial_{c} (\vert\omega\vert Y^{c}_a) = C^b_{ab}\,\vert\omega\vert\,. $

If we now apply the Killing conditions on the fields,
$\displaystyle {\bf Y}_a\Phi \rightarrow 0,\,\quad {\bf Y}_a{\bf Y}_b \Phi\rightarrow 0\,, $
we end up with 
$$\displaystyle \left(\frac{\delta{\mathcal L}}{\delta \Phi} \right)
       _{\!_{\! R}}
     =  \vert\omega\vert\bigg\{\frac{\delta{\mathcal L}_{R}}{\delta \Phi}
        \ - \ C_{ac}^c
          \left(\frac{\partial\tilde{\mathcal L}}{\partial{\bf Y}_a\Phi}\right)
            _{\!_{\! R}}
         \ + \ \frac{1}{2}  C_{ac}^c C_{bd}^d
        \left(\frac{\partial\tilde{\mathcal L}}{
            \partial{\bf Y}_a{\bf Y}_b\Phi} \right)
            _{\!_{\! R}}
    \bigg\}\,.$$

Since in general the pieces of the type 
$\left(\frac{\partial\tilde{\mathcal L}}{\partial{\bf Y}_a\Phi}\right)_{\!_{\! R}}$ will be different from zero, 
the last relation proves that  
\beq\displaystyle   \bigg(\frac{\delta{\mathcal L}}{\delta \Phi}
\bigg)_{\!_{\! R}}
        = \ \vert\omega\vert\bigg(\frac{\delta{\mathcal L}_{R}}{\delta \Phi}\bigg)
\Longleftrightarrow C_{ac}^c =0\,,\, \forall a 
\label{unim}
\eeq
which is the unimodularity condition mentioned in the introduction.
Abelian, semisimple, and compact groups are examples of groups fulfilling this condition.


\subsection{Reduction of the diffeomorphisms algebra}

Not all the elements of the gauge group will survive the reduction. 
The gauge group will be reduced to the elements
that act internally on the subset of solutions that share the Killing symmetries under which 
the reduction is taking place.
To show the general procedure it is enough to use a $1$-form ${\Omega}$ satisfying the Killing conditions, and
write it in the basis $\displaystyle\{{\bf d}x^\mu, {\omega}^{a}\}$,
$$\displaystyle {\Omega} = \Omega_\mu(x) {\bf d}x^\mu + \Omega_a(x)
(A^a_\mu(x) {\bf d}x^\mu + {\omega}^{a})\,.$$

The active diffeomorphisms expressed in this basis produce the change  
$$\displaystyle \Omega\rightarrow {\Omega'} = \Omega'_\mu(x,y) {\bf d}x^\mu + \Omega'_a(x,y)
(A'^a_\mu(x,y) {\bf d}x^\mu + {\omega}^{a})\,,$$
and if we require that the new object ${\Omega'}$ still satisfies the Killing conditions, that is, 
$$\displaystyle {\Omega'} = \Omega'_\mu(x) {\bf d}x^\mu + \Omega'_a(x)
(A'^a_\mu(x) {\bf d}x^\mu + {\omega}^{a})\,,$$
we will reduce the gauge group. In fact, not all diffeomorphisms will produce these
excusive $x$-dependences. Those that do are the diffeomorphisms that survive the reduction 
procedure, and define the reduced gauge group. 

\vspace{4mm}

Let us pause a moment to notice a somewhat subtle point. We are expressing
both the original objects --${\Omega}$- and the transformed ones --${\Omega'}$-- in the same
basis $\displaystyle\{{\bf d}x^\mu, {\omega}^{a}\}$. This fact is crucial for the results that 
follow and it is just a choice of a basis. The confusion may come from the fact that 
the ${\omega}^{a}$'s are themseves forms that change under an active diffeomorphism 
--active in the sense that moves the objects but not the coordinates-- and one may think of 
expressing ${\Omega'}$ in terms of $\displaystyle\{{\bf d}x^\mu, {\omega'}^{a}\}$. 
It is a matter of choice to proceed in one or another way, but the convenient choice for the 
practice of dimensional reductions is to stay always with the unique basis 
$\displaystyle\{{\bf d}x^\mu, {\omega}^{a}\}$\,. The reason is that the definition of the 
reduced Lagrangian (\ref{redlagr}) has been given precisely in terms of this unique basis. 

\vspace{4mm}

The diffeomorphisms belonging to this reduced gauge group turn out \cite{Pons:2003ka} 
to be of the form (for the infinitesimal generators),
$$\displaystyle  \vec v \rightarrow \epsilon^\mu(x)\partial_\mu 
+ \eta^a(x){\bf Y}_a + \xi^a(y){\bf
Y}_a\ $$
where $\displaystyle\epsilon^\mu(x)\partial_\mu\ $ generate diffeomorphisms in the reduced 
manifold;  \  
 $\displaystyle\eta^a(x){\bf Y}_a\ $ generate Yang-Mills transformations (and correspond to the 
inner automorphisms of the Lie algebra of Killing vectors); and finally \ 
$\displaystyle\xi^a(y){\bf Y}_a\ $ generate residual rigid symmetries 
(corresponding to outer automorphisms).

The transformations generated by the reduced diffeomorphisms are
\bea
\displaystyle\delta_{{}_{Diff}} g_{\mu\nu} &=& {\mathcal L}_{\vec\epsilon} g_{\mu\nu} 
\ \ {\rm (tensor)}\,,\nonumber \\
\displaystyle\delta_{{}_{Diff}} g_{ab} &=& {\mathcal L}_{\vec\epsilon} g_{ab}  
\ \ {\rm  (scalar)}\,,\nonumber \\
\displaystyle\delta_{{}_{Diff}} A^c_\mu &=& {\mathcal L}_{\vec\epsilon} A^c_\mu \ \ {\rm (vector)}\,.\nonumber
\eea

The Yang Mills gauge transformations are
\bea
\displaystyle\delta_{{}_{YM}} g_{\mu\nu} &=& 0\,,\nonumber \\
\displaystyle\delta_{{}_{YM}} g_{ab} &=& \eta^d
(C_{da}^c g_{cb} + C_{db}^c g_{ac})\,, \nonumber \\
\displaystyle\delta_{{}_{YM}}A^a_\mu &=&
\partial_\mu\eta^a +  A^c_\mu C_{cd}^a\eta^d\,, \nonumber \\
\displaystyle\delta_{{}_{YM}}\Omega_a  &=&\eta^d C_{da}^c
\Omega_{c} \,,\nonumber \\ 
\displaystyle\delta_{{}_{YM}}\Omega_\mu  &=& 0\,. \nonumber
\eea

And the residual rigid symmetries are 
\bea
\displaystyle\delta_{{}_{Res}} g_{\mu\nu} &=& 0\,, \nonumber \\
\delta_{{}_{Res}} g_{ab} &=&
-(B_a^c g_{cb} + B_b^c g_{ac})\,, \nonumber \\
\displaystyle\delta_{{}_{Res}}A^a_\mu &=&
 B^a_bA^b_\mu\,,  \nonumber \\
 \displaystyle\delta_{{}_{Res}} \Omega_a &=& - B_a^b \Omega_b ,\nonumber \\
{\rm with}&&\!\!\!\! 
\displaystyle C_{eb}^a B^e_{c} - C_{ec}^a B^e_{b} + C_{bc}^e B^a_{e} = 0 \,.\nonumber 
\eea

It is worth mentioning the appearance of these residual rigid symmetries, associated with the
outer automorphisms of the Lie algebra. In the abelian case these rigid symmetries define the 
general linear group $GL(n,R)$.

If our original manifold was $d+n$-dimensional and the quotient manifold 
--under the foliation produced by the $n$-dimensional algebra of Killing vectors-- 
is $d$-dimensional, 
the structure of the reduced gauge group  can be written as
$$\displaystyle Diff({\mathcal M}^{d+n}) \longrightarrow 
\left(Diff({\mathcal M'}^{d})\otimes {Res.}\right)\wedge
YM $$


\section{Second-type truncations: Constraints}

Introducing  constraints is the second way to reduce the degrees of freedom. Suppose that 
a dimensional reduction has been performed,  producing some scalar fields  $g_{ab}$ out 
of the higher dimensional metric. They transform under the adjoint action of the 
Yang-Mills gauge group, $\delta_{YM} g_{ab} = \eta^d (C_{da}^c g_{cb} + C_{db}^c g_{ac})$, 
which means that they are in general charged objects under YM. 
We can get rid of the charged scalars by imposing the constraint
\beq
\label{const}
C_{da}^c g_{cb} + C_{db}^c g_{ac} =0\,,
\eeq
but this ad hoc imposition entails consequences that we must control, as we shall see now. 
To be specific, consider that we have 
dimensionally reduced the Einstein-Hilbert  action  
$$S^{(d+n)} = \frac{1}{2\kappa^2} \int
d^dx\,d^ny\,{\vert{-\hat{g}_{\hat\mu\hat\nu}}\vert}^{1/2}
\,\hat{{\mathcal R}}\,,$$ 
from $d+n$ to $d$ dimensions, under a simple Lie algebra of $n$ independent Killing vector 
fields. The --consistently truncated-- 
reduced action becomes \cite{chofreund}\cite{Scherk:1979zr}
\bea
S^d &=& \frac{1}{2\kappa^2} \int d^d x\, \vert -
g_{\mu\nu}\vert^{1/2} \vert g_{ab}\vert^{1/2} \Bigg\{ {\mathcal
R}- \frac{1}{4} F^{\mu\nu a}\, F_{\mu\nu}^b\, g_{ab} 
 +\frac{1}{4} g^{\mu\nu}\, {\mathcal D}_\mu g_{ab}\, {\mathcal
D}_\nu g^{ab}\nonumber\\ 
&+& g^{\mu\nu}\, {\mathcal D}_\mu \ln \sqrt{\vert g_{ab}\vert} \,
{\mathcal D}_\nu \ln \sqrt{\vert g_{ab}\vert}
\displaystyle -\frac{1}{4}\,C^a_{bc}\left[ 2C^b_{a c^\prime} \, g^{c c^\prime} +
C^{a^\prime}_{b^\prime c^\prime}\, g_{aa^\prime}\, g^{bb^\prime}
\,g^{cc^\prime} \right] \Bigg\} \,.
\eea
 In this case the constraints (\ref{const}) amount to the survival of a single, neutral
scalar $\varphi$, and the constraints (\ref{const}) take the explicit form $g_{ab} = \varphi h_{ab}$ with 
$h_{ab}$ being
the Cartan-Killing metric. But imposing this substitution $g_{ab} = \varphi h_{ab}$ on the 
reduced lagrangian or on its reduced e.o.m. 
is not a commutative process in general --in the sense of the diagram  introduced in Section 2. 
In fact there is a mismatch in the equation for the 
neutral scalar $\varphi$ in the reduced theory and the original equations for the 
scalars $g_{ab}$. To overcome the mismatch, some more constraints are needed. In this 
case, the new constraints are 
\beq\label{constsec}
\frac{1}{4}F^{\mu\nu a}F_{\mu\nu}^{\ b}-
\frac{1}{4n}(F^{\mu\nu c}F_{\mu\nu}^{\ d}\, h_{cd}) h^{ab} = 0\,.
\eeq
The intepretation of these new constraints is straightforward within the framework of the 
Dirac-Bergmann theory of constrained systems \cite{ber1,ber2,ber3,dirac}. 
They are in fact the secondary constraints dynamically 
derived from the primary ones (\ref{const}). At this point it is necessary that we distinguish,
when considering second-type truncations, whether the constraints that are introduced preserve
or break the remaining gauge symmetry. In the second case we speak of gauge-fixing constraints.
Two theorems are worth considering in this respect. 
Suppose we start with a theory with Lagrangian ${\mathcal L}$ that gets reduced, 
by way of the introduction of some holonomic constraints, which we call ``primary'', 
to a reduced theory ${\mathcal L}_R$.

\noindent {\bf Theorem 1 (\it for non gauge-fixing constraints)}
{\it The theory ${\mathcal L}$ plus some added "primary" holonomic constraints is 
equivalent to the reduced theory ${\mathcal L}_R$ plus the "secondary" constraints dynamically 
inherited from the "primary" ones.}

This is what happenend in our example above. Indeed, a strong way to satisfy the secondary 
constraints (\ref{constsec}) is to set the YM vector potentials to zero. One can then show 
that no tertiary 
constraints appear, which means the we have ended up with a consistent truncation. The 
price, though, is that of having lost the YM structure. The proof of Theorem $1$ is 
given in \cite{Pons:2004ky}.

\noindent {\bf Theorem 2 (\it for gauge-fixing constraints)}
{\it The theory ${\mathcal L}$ plus some gauge-fixing constraints is
equivalent to the reduced
theory ${\mathcal L}_R$ plus the secondary constraints 
inherited from those primary constraints whose associated 
gauge symmetries are killed by the gauge-fixing.}

A good example of an application of this theorem is the Polyakov string in the conformal gauge, 
where the Virasoro constraints must be added by hand because the constraints introduced in the 
conformal gauge make them disappear in the reduced theory. A proof of this theorem is given in 
\cite{Pons:1995ss} for the case of total gauge-fixings and in Appendix C of 
\cite{Pons:1998tt} por partial ones.


\section{Reduction on coset spaces}

The first-type truncations considered so far do not include the very common case of dependent 
Killing vector fields, as happens in sphere reductions in general. The hard problem in this case is that 
there are no left invariant forms --under the group $G$-- on the coset $G/H$.

Let us first set some notation. We assume that the Lie algebra can be decomposed
$\displaystyle {\mathcal G} = {\mathcal H} + {\mathcal K}$ with 
$[{\mathcal H},{\mathcal H}]\in {\mathcal H}$ and
$[{\mathcal H},{\mathcal K}]\in {\mathcal K}$. 
In coordinates such that $y^a$ label the $H$-orbits (coordinates for the coset) and $z^i$
are coordinates along these orbits, a
generic element of $G$ is written $g(y^a, z^i)= L(y)h(z)$, where $h(z) \in H$ and 
$L(y)$ is a coset representative (see \cite{Salam:1981xd}).

For the group manifold $G$, 
the left invariant Lie algebra-valued Maurer-Cartan 
$1$-form $g^{-1} {\bf d}g$ is written as $g^{-1} {\bf d}g= \omega^a X_a + \omega^i X_i$. 
The vectors ${\bf K}_a,\   {\bf K}_i $ are defined as the dual vectors to the right invariant 
$1$-forms extracted from $ {\bf d}g\, g^{-1}$. 

Now for the coset. The Lie algebra-valued $1$-form
$ L^{-1} {\bf d}L = \theta^a X_a + \theta^i X_i$ allows to isolate the $1$-forms
$\theta^a = \omega^a_b(y,0){\bf d}y^b$, which are the closer we can get to the situation
with independent Killing vectors (group manifold) discussed in section 3. 
They may be used as a basis of $1$-forms 
for the coset space. The problem now is that, instead of being 
invariant under the left action of $G$, the forms $\theta^a$ satisfy
$${\mathcal L}_{{\bf K}_c}\theta^a = 
K^i_c(y,0)\,\omega^j_i(y,0)\,C^a_{jb}\,\theta^b \neq 0\,.
$$

We can now proceed to write the metric in a way similar to what we did before, 
$$ {\bf g} =  g_{\mu\nu} {\bf d}x^\mu {\bf d}x^\nu
        + g_{ab}\left(A^a_\mu {\bf d} x^\mu  +  {\theta}^{a}\right)
            \left(A^b_\nu {\bf d} x^\nu +  {\theta}^{b}\right)\,,$$
but it will not be easy to ensure ${\mathcal L}_{{\bf K}_c}{\bf g}=0$ becasue of the
lack of invariance of the $1$-forms $\theta^a$. Implementation of the
constraints  
$$C_{ia}^c g_{cb} + C_{ib}^c g_{ac} =0\,,\qquad A^b_\nu=0\,,$$  
will indeed guarantee that 
${\mathcal L}_{{\bf K}_c} {\bf g} =0 $. 
In this way one can get consistent truncations on coset spaces as in \cite{Karthauser:2006wb},
but the challenge remains to get consistent truncations without eliminating the YM gauge bosons.
To describe today's state of the art, let us quote the authors of \cite{Cvetic:2003jy}: 
{\it If one attempts a generalization of the reduction idea to a case where the internal 
manifold is a coset space, such as a sphere,
then aside from exceptional cases it is not possible to perform a 
consistent reduction that
retains a finite set of lower-dimensional fields including all the gauge bosons of the isometry
group. 
In those exceptional cases where such a consistent reduction is possible,
there is currently no clear understanding, for example from group theory, as to why the
consistency is achieved.}

\vspace{5mm}

All these difficulties notwithstanding, there are some 
impressive cases of consistent truncations in the literature. In fact they are truncations 
in the weaker sense mentioned in the 
introduction, but that does not make them less interesting. Let us mention the reduction of
$d=11$ supergravity to $d=4$ gauged $N=8$ supergravity on the coset space $S^7$ in the work of 
\cite{deWit:1986iy}; the reduction of
$d=11$ supergravity to $d=7$ gauged $N=2$ supergravity on the  coset space $S^4$ in
\cite{Nastase:1999cb,Nastase:1999kf}; the reduction of
$d=10$ type $IIB$ supergravity to $d=5$ gauged $N=8$ supergravity on the  coset space $S^5$ in 
\cite{Cvetic:2000nc}. Although not dealing with cosets, another interesting case worth 
mentioning is the reduction of 
$d=10$ $N=1$ supergravity to $d=4$ gauged $N=4$ supergravity on the group manifold 
$SU(2)\times U(1)^3$ in \cite{Chamseddine:1997nm,Chamseddine:1997mc}. The uplifting to $d=10$ 
of the remarkable solution found by these authors was interpreted in
\cite{Maldacena:2000yy}, in the context of the gauge/gravity 
correspondence, as a background dual to $d=4$, $N=1$ super Yang-Mills in the infrared.

\section*{Acknowledgments}
I would like to thank Larry Shepley and Pere Talavera for joint work related to some of 
the results presented here. I would like to thank also the organizers of the NEB XII 
conference (Nafplion, Greece, June-July 2006), specially Theodosios Christodulakis and 
Elias Vagenas, for their kind invitation and the excellent organization of the conference.
This work is supported in part by the European EC-RTN network MRTN-CT-2004-
005104, MCYT FPA 2004-04582-C02-01, CIRIT GC 2005SGR-00564.


\begin{thebibliography}{99}


\bibitem{Duff:hr}
M.~J.~Duff, B.~E.~Nilsson and C.~N.~Pope, ``Kaluza-Klein
Supergravity,'' Phys.\ Rept.\  {\bf 130} (1986) 1.


\bibitem{Appelquist:nr}
T.~.~Appelquist, A.~.~Chodos and P.~G.~Freund, ``Modern
Kaluza-Klein Theories,'' {\it  Reading, USA: Addison-Wesley (1987)
619 P. (Frontiers in Physics, 65)}.


\bibitem{gn} G. Nordstr\"om, Phys. Zeit. {\bf 15}, 504 (1914)

\bibitem{kaluza} 
T. Kaluza, ``On the problem of unity in physics'', Sitzunber. Preuss. Akad. Wiss. Berlin.
Math. Phys. {\bf K1}, 966 (1921).

\bibitem{klein} O. Klein, ``Quantum theory and 5-dimensional theory of relativity'', 
Z.F. Phyzik. {\bf 37}, 895 (1926); Nature {\bf 118} (1926)
516.

\bibitem{Duff:1994tn}
  M.~J.~Duff,
  ``Kaluza-Klein Theory In Perspective,''
  arXiv:hep-th/9410046.


\bibitem{pauli53} 
N. Straumann, ``On Pauli’s invention of non-Abelian Kaluza-Klein theory in 1953'' , grqc/
0012054;
L. O’Raifeartaigh and N. Straumann, ``Early history of gauge theories and Kaluza-Klein
theories, with a glance at recent developments'', hep-ph/9810524.

\bibitem{dewitt;63}
B.~S.~De Witt, In {\it C. and B.~S.~De Witt (ed.) ``Relativity,
Groups And Topology'', Gordan and Breach, New York (1964)}.

\bibitem{hawking;69}
S.~W.~Hawking, Mon.\ Not.\ R.\ Astron.\ Soc.\ {\bf 142} (1969)
129.

\bibitem{bianchi}
A. ~Krasinski, C. ~G. ~Behr, E. ~Schcking, F. ~B. ~Estabrook, H.
~D. ~Wahlquist, G. ~F. ~R. ~Ellis, R. ~Jantzen, W. ~Kundtpp.
``The Bianchi Classification in the Schcking-Behr Approach,''
Gen.\ Rel.\ Grav. {\bf 35} (2003) 475-489.

\bibitem{Scherk:1979zr}
J.~Scherk and J.~H.~Schwarz, ``How To Get Masses From Extra
Dimensions,'' Nucl.\ Phys.\ B {\bf 153} (1979) 61.


\bibitem{AlonsoAlberca:2003jq}
N.~Alonso Alberca, E.~Bergshoeff, U.~Gran, R.~Linares, T.~Ortin
and D.~Roest, ``Domain walls of D = 8 gauged supergravities and
their D = 11 origin,'' arXiv:hep-th/0303113.

\bibitem{Maccallum:gd}
M.~A.~MacCallum, ``Anisotropic And Inhomogeneous Relativistic
Cosmologies,'' In {\it   Hawking, S.W., Israel, W.: General
Relativity, an Einstein centenary survey. Cambridge University
Press, 1979, pags 533-580}.

\bibitem{Appelquist:1983vs}
T.~Appelquist and A.~Chodos, ``The Quantum Dynamics Of
Kaluza-Klein Theories,'' Phys.\ Rev.\ D {\bf 28} (1983) 772.

\bibitem{Palais1979}
R.~Palais,  Commun.\ Math.\ Phys.\ {\bf 69} (1979) 19-30.

\bibitem{Fels:2001rv}
  M.~E.~Fels and C.~G.~Torre,
  ``The principle of symmetric criticality in general relativity,''
  Class.\ Quant.\ Grav.\  {\bf 19} (2002) 641
  [arXiv:gr-qc/0108033].

\bibitem{Anderson:1999cn}
  I.~M.~Anderson, M.~E.~Fels and C.~G.~Torre,
  Commun.\ Math.\ Phys.\  {\bf 212} (2000) 653
  [arXiv:math-ph/9910015].

\bibitem{pope}
M.~Cvetic, G.~W.~Gibbons, H.~Lu and C.~N.~Pope,
``Consistent group and coset reductions of the bosonic string,''
arXiv:hep-th/0306043.

\bibitem{Pons:1998tt}
  J.~M.~Pons and L.~C.~Shepley,
   ``Dimensional reduction and gauge group reduction in Bianchi-type
  cosmology,''
  Phys.\ Rev.\ D {\bf 58} (1998) 024001
  [arXiv:gr-qc/9805030].

\bibitem{Pons:2003ka}
  J.~M.~Pons and P.~Talavera,
  ``Consistent and inconsistent truncations. General results and the issue  of
  the correct uplifting of solutions,''
  Nucl.\ Phys.\ B {\bf 678}, 427 (2004)
  [arXiv:hep-th/0309079].



\bibitem{chofreund}
Y.~M.~Cho and P.~G.~Freund,
``Nonabelian Gauge Fields In Nambu-Goldstone Fields,''
Phys.\ Rev.\ D {\bf 12} (1975) 1711.

\bibitem{ber1}
P.~G.~Bergmann (1949),
``Non-Linear Field Theories,'' Phys.\ Rev.\  {\bf 75} 680.\vspace{2mm}

\bibitem{ber2}
P.~G.~Bergmann and J. ~H. ~M. ~Brunings (1949),
`` Non-Linear Field Theories II. Canonical Equations and Quantization,''
Rev.\ Mod.\ Phys.  {\bf 21} 480.\vspace{2mm}

\bibitem{ber3}
J.~L.~Anderson and P.~G.~Bergmann (1951),
``Constraints In Covariant Field Theories,''
Phys.\ Rev.\  {\bf 83} 1018.\vspace{2mm}

\bibitem{dirac}
P.~A.~M.~Dirac (1950), ``Generalized Hamiltonian Dynamics,'' 
Can.\ J.\ Math.\  {\bf 2} 129-148.\vspace{2mm}


\bibitem{Pons:2004ky}
  J.~M.~Pons and P.~Talavera,
   ``Truncations driven by constraints: Consistency and conditions for  correct
  upliftings,''
  Nucl.\ Phys.\ B {\bf 703} (2004) 537
  [arXiv:hep-th/0401162].



\bibitem{Pons:1995ss}
  J.~M.~Pons,
  ``Plugging the gauge fixing into the Lagrangian,''
  Int.\ J.\ Mod.\ Phys.\ A {\bf 11} (1996) 975
  [arXiv:hep-th/9510044].



\bibitem{Salam:1981xd}
  A.~Salam and J.~A.~Strathdee,
  ``On Kaluza-Klein Theory,''
  Annals Phys.\  {\bf 141} (1982) 316.


\bibitem{Karthauser:2006wb}
  J.~L.~P.~Karthauser and P.~M.~Saffin,
  ``The dynamics of coset dimensional reduction,''
  Phys.\ Rev.\ D {\bf 73}, 084027 (2006)
  [arXiv:hep-th/0601230].

\bibitem{Cvetic:2003jy}
  M.~Cvetic, G.~W.~Gibbons, H.~Lu and C.~N.~Pope,
  ``Consistent group and coset reductions of the bosonic string,''
  Class.\ Quant.\ Grav.\  {\bf 20} (2003) 5161
  [arXiv:hep-th/0306043].

\bibitem{deWit:1986iy}
  B.~de Wit and H.~Nicolai,
  ``The consistency of the S**7 truncation in d = 11 supergravity,''
  Nucl.\ Phys.\ B {\bf 281} (1987) 211.



\bibitem{Nastase:1999cb}
  H.~Nastase, D.~Vaman and P.~van Nieuwenhuizen,
  ``Consistent nonlinear K K reduction of 11d supergravity on AdS(7) x S(4)
  and self-duality in odd dimensions,''
  Phys.\ Lett.\ B {\bf 469}, 96 (1999)
  [arXiv:hep-th/9905075].


\bibitem{Nastase:1999kf}
  H.~Nastase, D.~Vaman and P.~van Nieuwenhuizen,
  ``Consistency of the AdS(7) x S(4) reduction and the origin of  self-duality
  in odd dimensions,''
  Nucl.\ Phys.\ B {\bf 581}, 179 (2000)
  [arXiv:hep-th/9911238].
  
\bibitem{Cvetic:2000nc}
  M.~Cvetic, H.~Lu, C.~N.~Pope, A.~Sadrzadeh and T.~A.~Tran,
  ``Consistent SO(6) reduction of type IIB supergravity on S(5),''
  Nucl.\ Phys.\ B {\bf 586}, 275 (2000)
  [arXiv:hep-th/0003103].

\bibitem{Chamseddine:1997nm}
  A.~H.~Chamseddine and M.~S.~Volkov,
  ``Non-Abelian BPS monopoles in N = 4 gauged supergravity,''
  Phys.\ Rev.\ Lett.\  {\bf 79}, 3343 (1997)
  [arXiv:hep-th/9707176].

\bibitem{Chamseddine:1997mc}
  A.~H.~Chamseddine and M.~S.~Volkov,
  ``Non-Abelian solitons in N = 4 gauged supergravity and leading order  string
  theory,''
  Phys.\ Rev.\ D {\bf 57}, 6242 (1998)
  [arXiv:hep-th/9711181].

\bibitem{Maldacena:2000yy}
  J.~M.~Maldacena and C.~Nunez,
  ``Towards the large N limit of pure N = 1 super Yang Mills,''
  Phys.\ Rev.\ Lett.\  {\bf 86}, 588 (2001)
  [arXiv:hep-th/0008001].


\end{thebibliography}
\end{document}